# Analytic Method for Estimating Aircraft Fix Displacement from Gyroscope's Allan-Deviation Parameters

Jonathan. M. Wheeler, Jacob N. Chamoun, Vinayak Dangui, and Michel J. F. Digonnet

*Abstract*—The noise and drift requirements for a navigation-grade gyroscope are widely known, yet there is no simple analytic model of how the noise and drift of a gyroscope influence the fix displacement error (FDE) of an inertial navigation system (INS). This work derives simple analytical expressions for the cross-track and along-track errors of an aircraft whose INS consists solely of a three-axis gyroscope system with perfect knowledge of the vertical direction. The error signal of each gyroscope is Gaussian white noise and drift modeled as a first-order Markov random walk. These expressions provide a straightforward mean of calculating the FDE of an aircraft as a function of the flight duration, velocity, noise amplitude, drift amplitude, and drift's time constant. These expressions are validated with Monte-Carlo simulations of long flights. This model quantifies the noise-versus-drift trade-off for a gyroscope in an inertial navigation system. It can save time when estimating the noise and drift that a gyroscope must exhibit to satisfy a given position-error requirement, or vice versa. They are used in particular to confirm the values, often cited without demonstration, of the noise and drift required to meet the maximum position error of an aircraft imposed by the Federal Aviation Administration's required navigation performance 10 specification. Finally, it demonstrates that using the minimum in the measured Allan deviation of a gyroscope as a metric of the drift is incorrect, because it fails to capture the drift's time constant. The proper metric is the maximum in the Allan deviation.

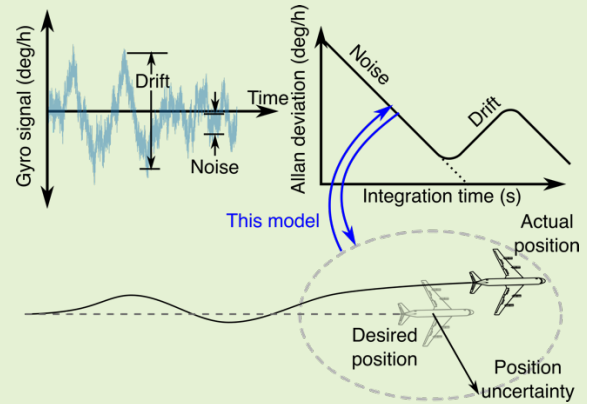

## I. Introduction

INERTIAL sensors such as gyroscopes and accelerometers are widely used for navigation of aircraft, ground vehicles, spacecraft, and submarine vehicles. These sensors are integrated in an inertial measurement unit (IMU) to provide real-time acceleration and angular rotation rate data [1]. An IMU is often incorporated into an inertial navigation system (INS) that integrates IMU acceleration and rotation-rate data to provide heading and position estimates [1]. A global navigation satellite system (GNSS) receiver can also be incorporated into an INS. However, GNSS signals may not always be available. In GNSS-denied environments, the uncertainties of the INS' estimates grow without bound in time.

The Federal Aviation Administration (FAA) standards divide an aircraft's fix displacement error (i.e., the distance between an INS' position estimate and the true position at a given point during a flight) into two axes. The along-track (ATRK) error is the displacement parallel to the intended flight trajectory (the *x* axis, see Fig. 1); the cross-track (XTRK) error is the displacement perpendicular to the trajectory (the *y* axis). For many trans-oceanic routes, the FAA requires an aircraft's INS to satisfy Required Navigation Performance 10 (RNP 10), which requires a 95%-confidence interval less than 10 nautical miles (nmi) for the geometric sum of the XTRK and ATRK errors [2]. For a 10-h long-distance flight, this represents a departure from actual position of 1 nmi/h. In GNSS-denied environments, rotation-rate and acceleration errors accumulate into uncertainty in the heading and velocity, respectively, and the fix displacement error statistically grows with time. Thus, the accuracy requirements are most challenging to meet at the end of a long flight.

Absent any external measurements of its position (e.g., GNSS or visual updates, star finding, magnetometers) an INS updates its position estimate with measurements from inertial sensors. Schuler tuning [3] can be used to bound errors in the

This work was supported by the Northrop Grumman Corporation.

J. M. Wheeler is with the Department of Electrical Engineering at Stanford University, Stanford, CA 94305 USA (e-mail: jamwheel@stanford.edu).

J. N. Chamoun was with the Department of Applied Physics at Stanford University, Stanford, CA 94305 USA. He is now with PARC, Palo Alto, CA, 94304, USA (jchamoun@parc.com).

V. Dangui was with the Department of Electrical Engineering at Stanford University, Stanford, CA 94305 USA. He is now with Facebook, Menlo Park, CA 94025, USA (vdangui@fb.com).

M. J. F. Digonnet is with the Department of Applied Physics at Stanford University, Stanford, CA 94305 USA (e-mail: silurian@stanford.edu).

accelerometers, but not the gyroscopes. Hence the gyroscope errors set the limit for the long-term performance of the INS.

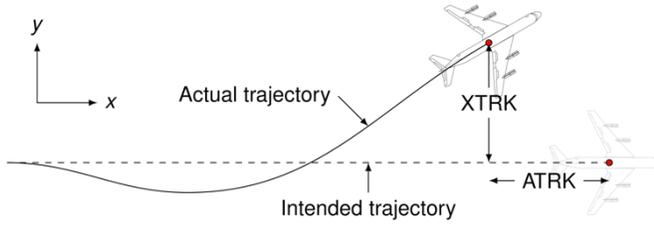

Fig 1. An example of how the fix displacement error is decomposed into XTRK and ATRK errors.

The literature states that a gyroscope meets the navigation-grade performance criterion if its bias drift is less than 0.01 deg/h [1], [4], [5], [6]. The origin of this number can be appreciated with the line of reasoning presented in [1], [4]: Consider an aircraft traveling along a great circle around the earth. To maintain constant attitude relative to the local horizon, it must push its nose down one degree for every 60 nautical miles of travel. Thus, one nautical mile of along-track displacement corresponds to a 0.016-degree change in pitch in the aircraft's frame of reference. In order for the ATRK error to grow slower than 1 nmi/h, the pitch gyroscope's angle uncertainty can grow no faster than 0.016 deg/h, in broad agreement with the value of 0.01 deg/h.

If the drift of the gyroscope evolves deterministically in time, it can be measured and subtracted from the 0.016-deg/h benchmark. However, the error of most gyroscopes is a combination of noise and drift, which are random processes, necessitating further analysis. Previous studies have presented analytical models of the position error due to noise and a constant bias offset in an INS [7], but in these studies the impact of drift (a time-dependent offset) (e.g., a first-order Markov random walk or autoregressive process) is estimated by Monte Carlo simulations. An analytic formulation would both be much faster and provide greater intuition into the impacts of the drift and noise on the position error caused by gyroscope drift than numerical simulations.

This work investigates the evolution of the along-track and cross-track errors with time in an aircraft guided by three axes of perfect accelerometers and three mutually orthogonal gyroscopes with known error statistics. Only the noise and drift in the gyroscopes are considered, since velocity errors in the accelerometers can be corrected (e.g., by Schuler tuning [3]). Closed-form expression are derived for the expected ATRK and XTRK errors, and for the fix displacement error, which is their geometric sum. These expressions provide for the first time a simple tool to calculate the fix displacement error of an aircraft as a function of the flight duration and velocity, and of the noise amplitude, drift amplitude, and drift time constant in the gyro output. These expressions are confirmed with Monte Carlo simulations of trans-Pacific flights. This simple analytical tool is then applied to demonstrate four important points. First, it confirms the often-cited values of the noise and drift that a gyroscope must have to meet inertial-navigation grade, namely 0.005 deg/h/√Hz and 0.01 deg/h, respectively. Second, it shows that these two values are not unique: over a small range of noise and drift values, a trade-off can be made between noise and drift, in that a gyro with a very small noise can tolerate a larger drift (or vice versa) and still be inertial grade. The analytical expressions spell out this relationship in simple terms. Third, this model shows that measuring the minimum in the Allan deviation to estimate the drift of a gyroscope is insufficient to determine the fix displacement error, and therefore whether a gyro satisfies RNP 10 or some other standard. A figure-of-merit that reflects the long-term drift statistics (e.g., the Allan deviation maximum) should be used instead. Fourth, the drift requirement is greatly relaxed for drift processes with a shorter time constant, and easy to quantify with this model. These equations readily lend themselves to trade-off studies to determine how changes in these critical parameters (flight velocity and total duration, and gyroscope noise and drift) affect the uncertainty in the aircraft's position at any time during a flight.

## II. MODEL OF THE NOISE AND DRIFT

The error sources in a gyroscope fall in two categories. The first one is errors that vary rapidly (on the order of fractions of a second or less). They are referred collectively as noise. They have a flat power-spectral density and dominate the signal at high frequencies. In a fiber optic gyroscope (FOG) for example, the noise contributions are typically a combination of photodetector noise, optical and electronic shot noise, laser relative intensity noise, and backscattering noise [8]. The second category is sources of error whose amplitudes vary slowly (seconds to hours or days) and undergo a bounded random walk. They are herein and generally referred to as drift and dominate the signal at low frequencies. Typical sources of drift in a FOG arise for example from asymmetric temperature fluctuations in the sensing fiber coil temperature (the Shupe effect [9]), the effect of Earth's magnetic field on the fiber index through the Faraday effect [10], [11] polarization coupling [12] and the Kerr effect [13] in the fiber coil, power fluctuations in the light source, and thermal drift in the electronics. All these contributions produce an offset that varies (drifts) over time. A gyroscope therefore outputs $\Omega_m[t]$, which is the sum of signal due to the applied rotation $\Delta\Omega_r[t]$ to be measured, the noise) $\Delta\Omega_N[t]$, and the drift $\Delta\Omega_K[t]$:

$$\Omega_m[t] = \Omega_r[t] + \Delta\Omega_N[t] + \Delta\Omega_K[t] \tag{1}$$

where square brackets represent indexing in discrete time, as would be done by an aircrafts' digital INS system.

The total noise in (1) is modeled as Gaussian white noise, expressed in deg/√h and described mathematically as

$$\Delta\Omega_N[t] = w_N / \sqrt{dt} \tag{2}$$

where $w_N$ is a random variable with a variance $N^2$, and $1/dt$ is the detection bandwidth. Square brackets denote discrete time, which becomes continuous in the limit of vanishingly small $dt$.

The variance of the noise $N^2$ is the sum of the variances of all the sources of noise (photodetector, shot, etc.).

Each of the drift contributions is a *rate* random walk process, which is modeled herein as a first-order Markov process [14], which also commonly termed a first-order autoregressive model [15], [16]. It models a $1/f^2$ random walk with an exponentially decaying memory, and produces a rotation-rate error $\Delta\Omega_K[t]$ that evolves over time according to [17], [18]

$$\Delta\Omega_K[t] = \Delta\Omega_K[t-dt]e^{-dt/T_c} + w_K\sqrt{dt} \quad (3)$$

where $w_K$ is a random variable with a variance $K^2$, and $T_c$ is the characteristic time constant of the process. $K$ and $T_c$ are determined by the physical nature of the source of drift. $T_c$ might be hours for drift due to laser-power fluctuations, and only minutes for drift caused by the Shupe effect. In this work, $T_c$ and $K$ are assumed to be constant, but in practice, these values could vary over the course of a flight.

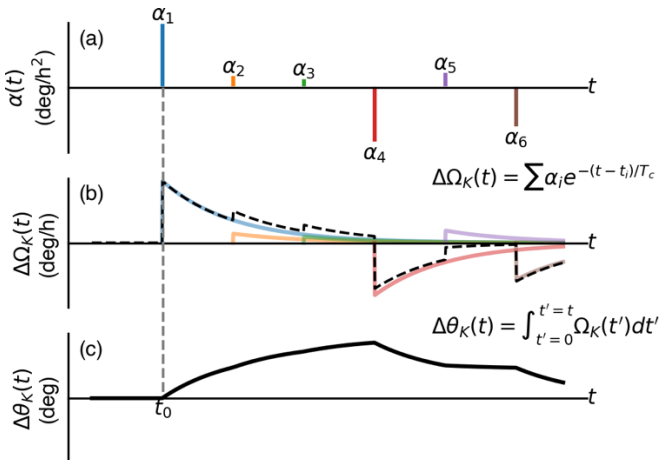

Fig. 2. (a) A series of random impulse signals $\alpha$ perturbs the gyroscope output signal rotation rate reading. (b) Each impulse produces an error signal in the rotation rate $\Delta\Omega_K$ that decays exponentially over time; the black trace is the total rotation-rate error signal. (c) The accumulated angular error $\Delta\theta_K$ is the integrated rotation-rate error.

This Markov process can be understood physically with reference to Fig. 2a. Every small time step $dt$, a new small perturbation $w_k$ is added to the existing rotation-rate error, for example due to a small change in local temperature of the fiber coil (giving rise to a small Shupe drift), or in power in the input light (giving rise to a small change in the gyro output signal). These sudden perturbations are represented in Fig. 2a by rotation-rate impulses of different colors, each with a random sign and amplitude. An impulse is represented mathematically by the second (driving) term in (2). The error $\Delta\Omega_K$ induced by an impulse decays exponentially with a time constant $T_c$. At any given time $t$, the rotation-rate error $\Delta\Omega_K[t]$ is then equal to $\Delta\Omega_K[t-dt]$, the value it had at $t-dt$, times $\exp(-dt/T_c)$ to account for this decay (the first term in (3)). In reality, these perturbations occur randomly in time. However, without loss of generality, they can be modeled as occurring periodically, which allows us to use (3).

Figure 2b illustrates the temporal evolution of $\Delta\Omega_K(t)$ due to a Markov process with a given $K$ and $T_c$ constants. Each impulse produces a contribution that decays exponentially with time (Fig. 2b). The total drift $\Delta\Omega_K$ is therefore the sum of these individual contributions (solid black curve in Fig. 2b). The angular heading error $\Delta\theta_K$ is then determined by summing (or integrating if $dt$ is small) the rotation-rate error $\Delta\Omega_K$ over time (solid curve in Fig. 2c). A histogram of a time trace of $\Delta\Omega_K$ over many $T_c$ would resemble a Gaussian distribution with a standard deviation of $K\sqrt{(T_c/2)}$. In other words, the error in $\Delta\Omega_K$ is more significant if the amplitude of the random steps is large, or if the decay-time is large.

Noise and drift are quantified in practice by measuring an Allan-deviation plot [19]. The measurement consists in recording the output of the gyroscope over an extended period, $T$ (hours or days) in the absence of rotation. This measurement produces a temporal trace $\Omega(t)$. The Allan variance is by definition [20] the expectation value of the difference between the means of two consecutive measured values $\Omega(t)$ calculated over an integration time $\tau$, and plotted as a function of $\tau$. The Allan variance is expressed mathematically,

$$\sigma^2(\tau) = \frac{1}{T-2\tau}\int_{t_0=0}^{t_0=T-2\tau}\left(\frac{1}{2\tau}\int_{t_0+\tau}^{t_0+2\tau}\Delta\Omega(t)dt - \frac{1}{2\tau}\int_{t_0}^{t_0+\tau}\Delta\Omega(t)dt\right)^2 dt_0$$
(4)

and the Allan deviation is $\sigma(\tau)$, the root of the Allan variance.

For any noise that is Gaussian white noise of amplitude $N$ as described by (2), and for any drift described by the Markov process of (3), it can be shown [19] that the Allan variance calculated by inserting (2) and (3) into (4) gives:

$$\sigma^2(\tau) = \sigma_N^2(\tau) + \sigma_K^2(\tau)$$
$$= \frac{N^2}{\tau} + \frac{K^2 T_c^2}{\tau}\left(1 - \frac{T_c}{2\tau}\left(3 - 4e^{-\frac{\tau}{T_c}} + e^{-\frac{2\tau}{T_c}}\right)\right) \quad (5)$$

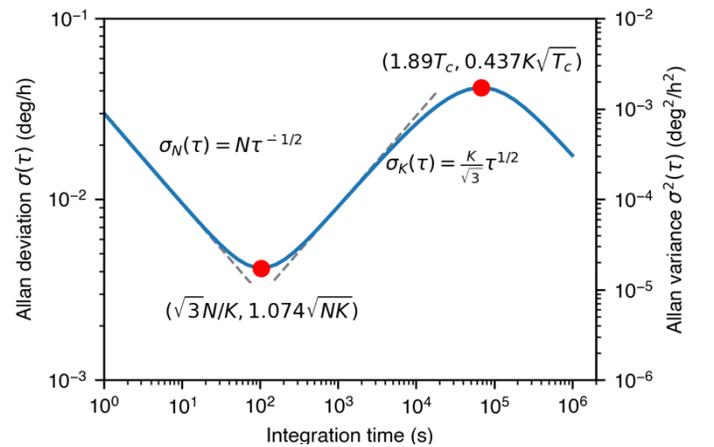

Fig. 3. An example Allan-deviation graph of a Gaussian white noise signal with $N = 0.03(°/h)/\sqrt{Hz}$, added to a first-order Markov process with $K = 0.03°/\sqrt{h^3}$ and $T_c = 10$ h. It exhibits a local minimum of $1.074\sqrt{NK}$, and a local maximum of $0.437K\sqrt{T_c}$.

In Figure 3 a generic Allan-deviation plot on a log-log scale illustrates these dependencies and the impact of noise and drift on this type of curve. The values of $N$, $K$, and $T_c$ used to generate it from (5) are listed in the figure. The noise processes (first term in (5)) create a line $\sigma_N(\tau) = N\tau^{-1/2}$ that starts at the top left of the figure and decreases with a slope of -1/2. The vertical position of this line is determined by the standard deviation (or amplitude) of the noise $N$. The drift (second term in (5)) appears at longer integration times as a line $\sigma_K(\tau) = K(\tau/3)^{1/2}$. It causes the Allan deviation to reach a minimum, then to rise with a slope +1/2. The vertical position of this line is determined by the standard deviation of the drift process $K$. At the right of Fig. 3, the Allan deviation does not increase indefinitely, because at some point the integration time $\tau$ exceeds the characteristic time constant $T_c$ of the random-walk process, which is therefore averaged out by the integration in (4), and the Allan deviation rolls off again with a slope of -1/2.

An Allan-deviation curve is often measured to characterize the noise and drift of a gyroscope. Specifically, the ordinate of the curve at an integration $\tau = 1$ s gives the ARW $N$ (in units (deg/h)/√Hz), while the ordinate of the first minimum is often used as a measure of the drift. For example, the Allan deviation of Fig. 3 shows that the value of the Allan deviation at 1 s is 0.03 deg/h, or $N = 0.03$ deg/h/√Hz, and the drift is the ordinate of the lower red point, or $4.1 \cdot 10^{-3}$ deg/h.

Although the derivation in the rest of this paper assumes only one drift process, its extension to multiple independent drifts is straightforward: Allan deviations with multiple sources of drift can be modeled with the addition of one or more Markov processes with individual $K_i$ and $T_{c,i}$ values.

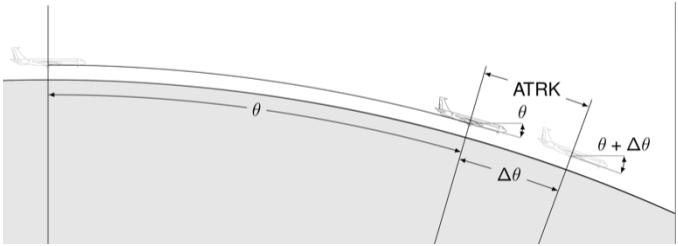

Fig. 4. The along-track distance of an aircraft is given by $R\theta$. The ATRK error is given by $R\Delta\theta$.

### III. MODELING THE ATRK ERROR

To estimate the ATRK error, it is assumed that the aircraft is level to its local horizon, and that the along-track displacement is inferred from its measured rotation. For a vehicle traveling along an arc on the surface of a sphere of radius $R$, the arc length is $R\theta$ (see Fig. 4). The along-track error is $R\Delta\theta$, where $\Delta\theta$ is the difference between the true angle it has subtended during the flight, and the gyroscopes' estimate of that value (see Fig. 4). The true angle is unknown but a deterministic quantity. The estimated angle is a measured quantity with some error attached to it due to the gyro's noise and drift: it is a statistical quantity. The statistics of $\Delta\Omega_K$ and $\Delta\Omega_N$ are used to determine the uncertainty of the integrated angle error $\Delta\theta$.

#### A. Along-Track Error Due to Noise

Consider the impact of the noise first. For a zero-mean Gaussian-white-noise rotation-rate error given by (2), the variance of the accumulated angle $d\theta_N$ during an arbitrarily short time interval $[t, t+dt]$ is

$$\sigma^2_{A,N}(d\theta_N(t)) = \left\langle (\Delta\Omega_N(t)dt)^2 \right\rangle = \left(\frac{N}{\sqrt{dt}}\right)^2 dt^2 = N^2 dt \quad (6)$$

Assuming that the measured noise values are statistically independent, their variances can be added. Integrating (6), namely the variance $\sigma^2_{A,N}$ of the angle $\Delta\theta_N$ accumulated over a period $t$, gives

$$\sigma^2_{A,N}(\Delta\theta_N(t)) = \int_{t'=0}^{t'=t} \sigma^2_{A,N}(d\theta_N(t')) = \int_{t'=0}^{t'=t} N^2 dt' = N^2 t \quad (7)$$

#### B. Along-Track Error Due to Drift

For the drift, the error-accumulation process described in Fig. 2 can be described mathematically as follows. At some time $t_0$ during the flight, an impulse error $\alpha(t_0)$ occurs (blue impulse in Fig. 2a). This impulse induces a small error in the rotation rate reading $\Delta\Omega(t)$ that decays over time with a time constant (blue trace in Fig. 2b). In turn, this error gets integrated over time by the gyro readout system into a cumulative angle (or heading) error $\Delta\theta(t)$ that grows to a finite asymptotic value at times $t \gg T_c$ (start of black trace in Fig. 2c). Since the distance traveled on the surface of a sphere of radius $R$ is $R\Delta\theta$, the cumulative ATRK error expressed as a distance is $\Delta\theta(t)$ multiplied by $R$ (not shown in Fig. 2). The along-track error at time $t$ caused by a single impulse error at time $t_0$ of amplitude $w_K = \alpha(t_0)$ can then be expressed as

$$\Delta\Omega_{A,K}(t,t_0) = \alpha(t_0) e^{-\frac{(t-t_0)}{T_c}} \quad (8)$$

The total angular error at time $t$ caused by the sum of the impulses occurring in some earlier interval $[t_0, t_0+dt]$ is then

$$\sigma^2_{A,K}(d\theta_K(t,t_0)) = \left\langle \left( \int_{t'=t_0}^{t'=t_0+dt} \alpha(t')dt' \times \int_{t'=t_0}^{t'=t} e^{-\frac{(t'-t_0)}{T_c}} dt' \right)^2 \right\rangle$$
$$= K^2 T_c^2 \left( 1 - e^{-\frac{(t-t_0)}{T_c}} \right)^2 dt \quad (9)$$

where the first integral determines the net impulse in the interval $[t_0, t_0+dt]$, and the second integral propagates the residual effect of that impulse from $t_0$ up to $t$.

Assuming for now that the error in the gyro signal at the start of the flight ($t = 0$) is negligible (this contribution is treated in the next sub-section), the total ATRK error due to sources of drift that take place between $t = 0$ and the elapsed flight time at $t$ can be determined by integrating (9) for all the impulses occurring in the interval $[0, t]$. This integration gives the variance of the total accumulated angular error at time $t$

$$\sigma_{A,K}^2(\Delta\theta_K(t)) = \int_{t_0=0^+}^{t_0=t} \sigma_{A,K}^2(d\theta_K(t,t_0))$$
$$= K^2 T_c^2 \left(t - \frac{T_c}{2}\left(e^{-\frac{2t}{T_c}} - 4e^{-\frac{t}{T_c}} + 3\right)\right) \quad (10)$$

where $0^+$ means that the initial value of the error signal at $t = 0$ is treated as a special case and is not included in the integral.

### C. Along-Track Error Due to the Turn-on Error

Consider now the impact on the along-track error of the offset that exists in the gyroscope at the time it is turned on at the start of a flight. Because the effect of the impulse on the rotation drift $\Delta\Omega_K$ diminishes (exponentially) towards but never quite reaches zero, $\Delta\Omega_K$ can be considered to have an "infinite" memory of $\alpha$. In other words, at least some of the sources of drift that occurred prior to shutting off the gyroscope may not have fully decayed to zero when the gyroscope is brought back online and may therefore contribute to a turn-on error. This error is not constant, but decays over time, possibly over a long period of time if it has a long $T_c$, so it cannot simply be measured and subtracted.

The turn-on rotation-rate error is the accumulation of the integrated $\alpha$ signal for all times $t < 0$. The variance of $\Delta\Omega(t = 0)$ is then

$$\sigma_{A,0}^2(\Delta\Omega(t=0)) = \left\langle \left(\int_{t'=-\infty}^{t'=0} \alpha(t')e^{-t'/T_c} dt'\right)^2 \right\rangle = \frac{K^2 T_c}{2} \quad (11)$$

The variance of the heading error that this turn-on offset produces after time $t$ is

$$\sigma_{A,0}^2(d\theta_K(t,0)) = \sigma_{A,0}^2(\Delta\Omega(t=0))\left(\int_0^t e^{-\frac{t'}{T_c}} dt'\right)^2$$
$$= \frac{K^2 T_c^3}{2}\left(1 - e^{-\frac{t}{T_c}}\right)^2 \quad (12)$$

In some cases, this turn-on error (12) is negligible compared to the in-flight drift errors (given by (10)) because during the flight they are exponentially attenuated, whereas in-flight impulse errors keep occurring and build up. This is indeed easy to verify by comparing (11) and (9) when the total flight time $T$ is much larger than $T_c$. The ratio of these two equations is then $\sim T_c/(2T)$, and very small. For example, for a 10-hour flight, $T = 10$ h and $T_c = 30$ min, the error due to turn-on offset is 2.5% of the total drift error. It can be deemed negligible.

### D. Total Along-Track Error

As explained earlier, the variance of the total ATRK error (a distance) is $R^2$ multiplied by the sum of the variances of the angular heading (7), (8), and (12), or

$$\sigma_A^2(t) = N^2 R^2 t + K^2 T_c^2 R^2 \left[t - T_c\left(1 - e^{-t/T_c}\right)\right] \quad (13)$$

where the first term is due to error from the noise and the second term is due to drift, including the (small) error at turn-on.

## IV. MODELING THE XTRK ERROR

To estimate the XTRK, consider a vehicle moving in a straight line on a Cartesian plane. At any time $t$, it has a velocity $v(t)$ and angular heading error $\Delta\theta(t)$. For small heading errors, its distance from its intended trajectory grows as

$$dy(t) = v(t)\sin(\Delta\theta(t))dt \approx v(t)\Delta\theta(t)dt \quad (14)$$

The total XTRK error $y(t)$ at time $t$ is the integral of (14) over time, and the analytical expression for the XTRK error is the expectation value of the root mean-squared value of $y(t)$.

We assume the velocity is inferred from the time derivative pitch gyroscope's estimation of along-track displacement $x(t)$ (i.e., $v = dx(t)/dt = Rd\theta(t)/dt$). The pitch gyroscope will introduce a small error in velocity (1 km/h for an error of 0.01 °/h), but this velocity error is usually negligible (<1%) compared to the total speed of the aircraft (hundreds of km/h) and can be ignored in this analysis.

### A. Noise Contribution to XTRK Error

For a Gaussian-white-noise process, at each time $t_0$, the rotation-rate error is perturbed by $\Delta\Omega_N(t) = w_N/\sqrt{dt}$ (see (1)), and the angular heading by $\Delta\Omega_N(t)dt = w_N\sqrt{dt}$. If the velocity is constant, at some later time $t$ it causes a XTRK error of $vw_N(t-t_0)\sqrt{dt}$. The variance of this quantity is

$$\sigma_{X,N}^2(t,t_0) = v^2 N^2(t-t_0)^2 dt \quad (15)$$

Adding the variances of the contributions of all time $t'$ between 0 and $t$ gives a total variance due to the noise alone of

$$\sigma_{X,N}^2(t) = N^2 v^2 \int_0^t (t-t_0)^2 dt_0 = \frac{1}{3}N^2 v^2 t^3 \quad (16)$$

### B. Drift Contribution to XTRK Error

To determine the contribution of the first-order Markov process to the total XTRK error, the rotation-rate error is modeled in the same fashion as done in Section IIIB. The mathematical process is exactly the same as described in Fig. 2a–2c to obtain the cumulative heading error $\Delta\theta(t)$. The cross-track error is simply the integral over time of this heading error times the aircraft velocity $v$. The cross-track error (in terms of distance from the intended trajectory) caused by a single impulse error $\alpha(t_0)$ at time $t_0$ is then given by

$$y(t,t_0) = v\alpha(t_0)\int_{t_\theta=t_0}^{t_\theta=t}\int_{t_\Omega=t_0}^{t_\Omega=t_\theta} e^{-\frac{t_\Omega-t_0}{T_c}} dt_\Omega\, dt_\theta \quad (17)$$

the variance of which is

$$\sigma_{X,K}^2(t,t_0) = K^2 v^2 dt_0 \left( \int_{t_\theta=t_0}^{t_\theta=t} \int_{t_\Omega=t_0}^{t_\Omega=t_\theta} e^{-\frac{t_\Omega-t_0}{T_c}} dt_\Omega dt_\theta \right)^2 \quad (18)$$

where $t_\Omega$ is the integration variable for the integration of the rotation rate $\Omega$ into a heading $\theta$, and and $t_\theta$ for the integration of the heading into a change in a XTRK-position $y$.

Integrating over all impulse times $t_0$ over the duration of the flight $[0, t]$ yields

$$\sigma_{X,K}^2(t) = K^2 v^2 \int_{t_0=0}^{t_0=t} \left( \int_{t_\theta=t_0}^{t_\theta=t} \int_{t_\Omega=t_0}^{t_\Omega=t_\theta} e^{-\frac{t_\Omega-t_0}{T_c}} dt_\Omega dt_\theta \right)^2 dt_0$$

$$= K^2 T_c^2 v^2 \left[ \frac{t^3}{3} - t^2 T_c + t T_c^2 \left( 1 - 2e^{-\frac{t}{T_c}} \right) + \frac{T_c^3}{2} \left( 1 - e^{-\frac{2t}{T_c}} \right) \right] \quad (19)$$

### C. Turn-On Offset Contribution to XTRK Error

When the INS is turned on at the beginning of a flight $t = 0$, it has accumulated a finite rotation-rate error whose variance is given by (10). This error can in principle be subtracted, but assuming it is not, it will decrese over time during the flight, but it will also integrate into an angular error, and a cross-track positional error. The variance of this error can be calculated in the same manner as (19), with the difference that prior to $t = 0$ the gyroscope was not moving and the accumulated heading error is 0 at the beginning of the flight, so that the integrals now start at $t_\theta = 0$ and $t_\Omega = 0$. The variance of the XTRK error induced by this turn-on bias is then

$$\sigma_{X,K}^2(t,0) = K^2 v \int_{t_0=-\infty}^{t_0=0} \left( \int_{t_\theta=0}^{t_\theta=t} \int_{t_\Omega=0}^{t_\Omega=t_\theta} e^{-\frac{t_\Omega-t_0}{T_c}} dt_\Omega dt_\theta \right)^2 dt_0$$

$$= \frac{K^2 T_c^3 v^2}{2} \left( t - T_c \left( 1 - e^{-t/T_c} \right) \right)^2 \quad (20)$$

Here too, it is easy to see that after a long flight the turn-on error (20) can be negligible compared to the in-flight drift error (19) if $T \gg T_c$. In this limit, the ratio of these errors is then $\sim 3T_c/2T$, or $\sim 7.5\%$ when $T = 20T_c$.

### D. Total XTRK Error

Adding the variances from the noise process and the in-flight and turn-on drift processes results in the following expression for the total XTRK error at time $t$:

$$\sigma_X^2(t) = \frac{N^2 v^2 t^3}{3}$$
$$+ K^2 T_c^2 v^2 \left[ \frac{t^3}{3} - t^2 T_c + t T_c^2 (1 - 2e^{\frac{-t}{T_c}}) + \frac{T_c^3}{2} (1 - e^{-\frac{2t}{T_c}}) \right] \quad (21)$$
$$+ \frac{K^2 T_c^2 v^2}{2} \left( t - T_c \left( 1 - e^{-t/T_c} \right) \right)^2$$

$$(21)$$

where the first term is the error from the noise, the second term is the error from the drift during the trajectory, and the final term is the error from the offset present at turn-on.

(21) shows that the first two terms, which dominate as explained earlier, both grow monotonically with time. At times $t$ longer than $T_c$ (a few hours if $T_c$ is 30 min or less), they grow approximately as $t^3$, which is very fast. Shorter flights and shorter drift time constants will greatly reduce the positional error, in a predictable manner.

Together, (13) and (21) provide the average ATRK and XTRK rms errors for an aircraft that has been navigating for a period $t$ without a GNSS update. These errors are imposed by the gyroscopes only. They can be improved with sensor-fusion algorithms with other sensors (e.g., accelerometers, magnetometers, and pitot tubes). [21]

## V. VALIDATION USING MONTE CARLO SIMULATIONS

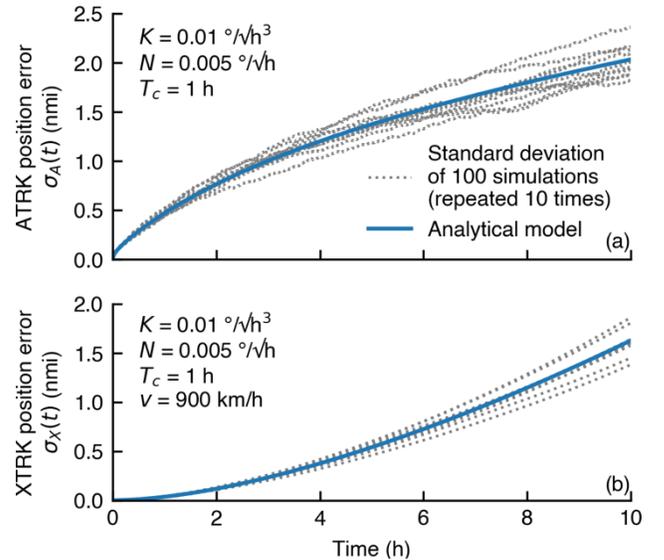

Fig. 5. The top shows the results for the ATRK component, and the bottom shows the results for the XTRK component.

To validate these derivations, we used Monte Carlo simulations to model the ATRK and XTRK errors of an aircraft traveling along a great circle at 900 km/h. The gyroscope was assumed to have an error signal determined by a Gaussian random noise and a single first-order Markov random walk. A temporal trace of the angular position of the aircraft was generated, in which at each time $t$ the deterministic position of the aircraft (given simply by its known velocity) was perturbed by a small amount of noise and a small amount of drift. These perturbations were chosen by a random-number generator so that they have a standard deviation $N$ and satisfy (2) for the noise, and a standard deviation $K$ and satisfy (3) for the drift. The estimates for the standard deviations of ATRK and XTRK errors were calculated from ensembles of 100 flights. Ten groups of 100 flights were simulated, resulting in ten ATRK and XTRK estimates at each time.

The standard deviations predicted by these Monte Carlo simulations are shown in dotted gray curves in Fig. 5. The analytical solutions for the ATRK error (12) and the XTRK error (20) are also shown as dark blue lines. For both errors the Monte Carlo simulation results agree well with the analytical results. A small sample size of 100 was chosen because larger sample sizes matched the analytic results too well and could not be distinguished from each other in Fig. 5.

## VI. Determining the Performance Requirements for Aircraft-Grade Gyros

The total fix displacement error (FDE) is defined as the sum of the ATRK and XTRK variances, $\sigma_{FDE}^2 = \sigma_A^2 + \sigma_X^2$. For a gyroscope to meet the requirements for aircraft navigation, the published standard benchmarks for noise and drift are 0.005 °/√h and 0.01 °/h, respectively, these values representing one standard deviation [4]. To validate these values, the analytical expressions of the variances of the ATRK (12) and XTRK errors (21) were used to calculate $\sigma_{FDE}^2$ for various values of $N$ and $K$ and a drift time constant $T_c = 1$ h. These values were calculated at the end of a 10-h flight at 900 km/h, or a total arc length around the Earth of 9000 km.

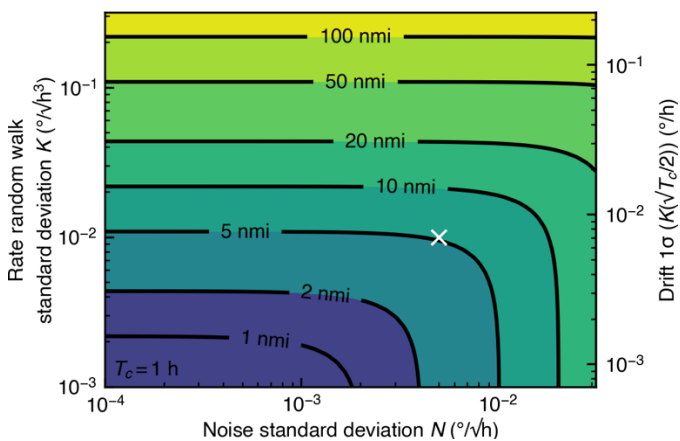

Fig. 6. The 95%-confidence interval of the total FDE after a 10-h flight along a great arc at 900 km/h. The rate random walk (drift) is quantified by the value of $K$, and the ARW (noise) by $N$.

Figure 6 plots the FDE's standard deviation $\sigma_{FDE}$ predicted by these simulations, represented in a 2D heat map as a function of $N$ and $K$. To represent the 95%-confidence interval, the quantity that is plotted is not $\sigma_{FDE}$ but $2\sigma_{FDE}$. The contour lines are lines of constant FDE. For example, on the contour label 10 nmi, the standard deviation of the FDE is $\sigma_{FDE} = 5$ nmi. It means that if one were to measure with a gyroscope whose $N$ and $K$ are on this curve the positional error for a large number of flights, the resulting probability distribution would be a Gaussian with a standard deviation $\sigma_{FDE} = 5$ nmi, so that 95% of the data points fall within $2\sigma_{FDE} = 10$ nmi of the center. The right axis in Fig. 6 shows the scale for the gyroscope drift in degrees/hour.

With reference to Fig. 6, as either $N$ or $K$ increases the FDE increases, as expected. For a given level of drift $K$, as the noise $N$ increases from a very small value the FDE remains constant at first, then it increases rapidly. This statement is also true when $N$ and $K$ are switched. The contours of Fig. 6 confirm the performance benchmarks mentioned above. To wit, consider a gyro with a noise $N = 0.005°/\sqrt{h}$ and a drift with a standard deviation $K = 0.01°/h^{3/2}$. For the assumed value of $T_c = 1$ h, and for these values of $N$ and $K$ (plotted as the white cross in Fig. 6, the 95% confidence interval of the FDE is just under 10 nmi, consistent with the RNP 10 requirement.

The shape of the contours in the logarithmic scale of Fig. 6 is close to a sharp corner, which fortuitously simplifies the interpretation of the FDE dependence on $N$ and $K$. Taking the 10-nmi contour as an example, for any value of the noise up to $\sim 10^{-2}$ deg/√h the drift needs to have a value of $K = 2.1 \cdot 10^{-2}$ deg/h$^{3/2}$ for the FDE to be 10 nmi. When the noise is higher than $\sim 10^{-2}$ deg/√h, the drift must be much smaller for RNP 10 to be satisfied. In fact, for a noise of $\sim 2 \cdot 10^{-2}$ deg/√h the drift must be zero. For higher noise values RNP 10 cannot be satisfied, the reason being that the noise contribution to the FDE alone exceeds 10 nmi. There is a narrow region of values of $N$ ($\sim 1$–$2 \cdot 10^{-2}$ deg/√h) and $K$ ($1.1 \cdot 10^{-2}$–$2.1 \cdot 10^{-2}$ deg/h$^{3/2}$) in which noise can be traded for drift, or vice versa.

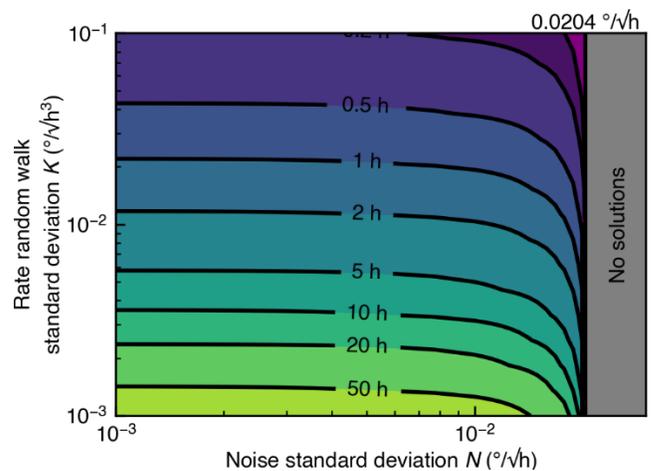

Fig. 7. Contours of the values of the drift time constant $T_c$ that produces a fix displacement error of 10 nmi as a function of the rate random walk constant $K$ and angular random walk $N$ for a 10-h flight.

The importance of the drift time constant $T_c$ is illustrated in Fig. 7, which plots in the ($N$, $K$) plane the values of $T_c$ required to meet the RNP-10 requirements. Along each contour, the coordinates define the values of the standard deviations of the rotation-rate noise and drift required for the FDE to be 10 nmi after a 10-h flight. The contours have the same general shape as in Fig. 6: (1) for a given $T_c$, when the noise is low the drift requirement is constant, and vice versa; and (2) there is only a small region where drift can be traded for noise. In the range of low noise typical of a high-accuracy gyroscope, namely under $10^{-3}$ °/√h, the drift requirement is greatly relaxed for drift processes with a shorter time constant. This dependence is sub-linear: when $T_c$ decreases from 10 h to 1 h, the drift requirement is relaxed from $\sim 3.7 \cdot 10^{-3}$ to $\sim 1.2 \cdot 10^{-2}$ °/h$^{3/2}$, a factor of only $\sim 3.2$. To be complete, this discussion must also acknowledge that most drift processes are bounded: the drift does not grow indefinitely but tends to oscillate between fixed, finite bounds. The standard deviation of the time trace produced by (3) is $\sigma_K = K(T_c/2)^{1/2}$, is therefore also bounded. This behavior implies that, given a fixed standard deviation $\sigma_K$, a process with a long $T_c$ will in general exhibit a smaller impulse amplitude $K$. So even though Fig. 7 states that a drift with a longer $T_c$ requires a lower $K$ value, it turns out that for a given $\sigma_K$, a drift with a longer $T_c$ does tend to have a lower $K$.

As described in relation to Fig. 4, the minimum in the Allan deviation is often used (without rigorous justification) as a measure of the drift of a gyro [3], [13]. The analysis presented here shows that the Allan deviation minimum is not appropriate to estimate the FDE. The reason can be gleaned from the expressions of the ATRK (13) and XTRK (21) errors. In both expressions, evaluation of the drift contribution requires knowledge of both $K$ and $T_c$. With reference to the typical Allan deviation of Fig. 3, the coordinates of the minimum in the Allan deviation depend on $K$ but not on $T_c$. A measurement of this minimum is therefore insufficient to calculate the FDE, and therefore to determine whether a gyroscope meets a particular requirement, for example for aircraft navigation. The Allan deviation minimum is not a suitable metric for the drift-induced heading or position errors. It merely represents the time scale at which drift processes become the dominant source of error.

A measurement of the maximum in the Allan deviation curve, on the other hand, will provide the necessary information. As shown in Fig. 4, the abscissa of this point is $1.89 T_c$, and its ordinate $0.437 K T_c^{1/2}$. The values of $K$ and $T_c$, and therefore of the drift contribution to the FDE, can then be calculated unambiguously from a measurement of the maximum. However, as is well known in the field, such a measurement is time consuming for long time constants $T_c$, requiring up to many days of continuous data collection.

## VII. CONCLUSION

The ATRK and XTRK errors of an INS guided only with gyroscopes traveling on the surface of a fixed-radius sphere have been analytically derived. The gyroscope error signal is modeled as an error signal composed of noise, described as Gaussian white noise (standard deviation $N$), and drift, represented by a first-order Markov random walk process (standard deviation $K$ and time constant $T_c$). The outcome of this derivation is a simple expression for the ATRK and XTRK errors as a function these three parameters, as well as the duration and velocity of the flight. This expression, confirmed with Monte-Carlo simulations of multiple-hour flights, can be used to quickly calculate the error in the location of an aircraft knowing these parameters. This model confirms the commonly accepted benchmark values found in the literature for the noise and drift required of a gyroscope to meet the FAA's RNP-10 requirement (an error in the knowledge of the aircraft position smaller than 10 nautical miles at all times during a flight). The model additionally shows that there is a small range of noise and drift values over which a trade-off can be made between noise and drift. It also shows that the drift requirement is stricter for drift processes with a long time constant. Finally, it demonstrates that measuring the minimum in the Allan deviation of the output of a gyro is insufficient to determine the fix displacement error; the Allan deviation maximum should be used instead. These closed-form analytical expressions can be used by system engineers performing trade studies as an alternative method of determining mission-worthiness of a set of Allan-deviation parameters, augmenting current methods of field-testing of the unit or Monte-Carlo system simulations.

## REFERENCES

[1] D. Titterton and J. Weston, *Strapdown Inertial Technology*, 2nd ed. Reston, VA: American Institute of Aeronautics, 2004, pp. 341, 544.
[2] Federal Aviation Administration, "Required navigation performance 10 (RNP 10) operational authorization" Nov. 2011. [Online]. Available: https://fsims.faa.gov/wdocs/orders/8400_12.htm
[3] M. Schuler, "Die Störung von Pendel und Kreiselapparatendurch die Beschleunigung des Fahrzeuges," *Physik. Zeitschr.*, vol. 24, no. 16, pp. 344–350, Aug. 1923.
[4] H. Lefèvre, *The Fiber-Optic Gyroscopes*, 2nd ed. Boston, MA: Artech House, 2014, pp. 26, 375–376, Figure 2.18.
[5] O. Deppe, G. Dorner, S. König, T. Martin, S. Voigt, and S. Zimmermann, "MEMS and FOG technologies for tactical and navigation grade inertial sensors—recent improvements and comparison," *Sensors*, vol. 17, no. 3, pp. 567–589, Mar. 2017.
[6] P. D. Groves, *Principles of GNSS, Inertial, and Multisensor Integrated Navigation Systems*, 2nd ed. Boston, MA: Artech House, 2013, pp. 208.
[7] W. S. Flenniken, "Modeling inertial measurement units and analyzing the effect of their errors in navigation applications." Master's thesis, Mech. Eng. Dept., Auburn University, Auburn, pp. 31–60, 2005.
[8] S. W. Lloyd, M. J. F. Digonnet, and S. Fan, "Modeling coherent backscattering errors in fiber optic gyroscopes for sources of arbitrary line width," *J. Lightwave Technol.*, vol. 31, no. 13, pp. 2070–2078, Jul. 2013.
[9] D. M. Shupe. "Thermally induced nonreciprocity in the fiber-optic interferometer," *Appl. Opt.* vol. 19, no. 5, pp. 654–655, Mar. 1980.
[10] K. Böhm, K. Petermann, and E. Weidel, "Sensitivity of a fiber-gyroscope to environmental magnetic fields," *Opt. Lett.*, vol. 7, no. 4, pp. 180–182, Apr. 1982.
[11] K. Hotate and K. Tabe, "Drift of an optical fiber gyroscope caused by the Faraday effect: influence of the earth's magnetic field," *Appl. Opt.*, vol. 25, no. 7, pp. 1086–1092, Apr. 1986.
[12] J. N. Chamoun and M. J. F. Digonnet, "Noise and bias error due to polarization coupling in a fiber optic gyroscope," *J. Lightwave Technol.*, vol. 33, no. 13, pp. 2839–2847, Apr. 2015.
[13] R. A. Bergh, H. C. Lefèvre, and H. J. Shaw, "Compensation of the optical Kerr effect in fiber-optic gyroscopes," *Opt. Lett.*, vol. 7, no. 6, pp. 282–284, June 1982.
[14] S. Nassar, K.-P. Schwarz, and N. El-Sheimy, "Modeling inertial sensor errors using autoregressive (AR) models," *Journal of the Institute of Navigation*, vol. 51, no. 4, pp. 259-268, Aug. 2014.
[15] M. Narasimhappa, J. Nayak, M. H. Terra, and S. L. Sabat, "ARMA model based adaptive unscented fading Kalman filter for reducing drift of fiber optic gyroscope," *Sens. Actuators, A,* vol. 251, no. pp. 42–51, Nov. 2016.
[16] J. Sun, X. Xu, Y. Liu, T. Zhang, and Y. Li, "FOG random drift signal denoising based on the improved AR model and modified Sage-Husa adaptive Kalman filter," *Sensors,* vol. 16 no. 7 pp. 1073, Jul. 2016.
[17] P. Iv, J. Lai, J. Liu, and G. Qin, "Stochastic error simulation method of fiber optic gyros based on performance indicators," *J. Franklin Inst.*, vol. 351, no. 3, pp. 1501–1516, Mar. 2014.
[18] C. J. Geyer, "Introduction to Markov Chain Monte Carlo," in *Handbook of Markov Chain Monte Carlo*, S. Brooks, A. Gelman, G. L. Jones, X.-L. Meng, Eds., New York City, NY: CRC Press, 2011, pp. 9–10.
[19] IEEE Aerospace and Electronic Systems Society. "IEEE standard specification format guide and test procedure for single-axis laser gyros," Sep. 2006.
[20] D. Allan, "Statistics of atomic frequency standards," *Proc. of IEEE*, vol. 54, no. 2, 221–230, Feb. 1966.
[21] R. E. Kalman, "A new approach to linear filtering and prediction problems." *Transactions of the ASME* vol. 82, no. 1, pp. 35–45, Mar. 1960.